\documentclass[jcp,amsmath,amssymb,showpacs, jcp, twocolumn]{revtex4-1}

\usepackage{hyperref}
\usepackage{graphicx}
\usepackage{bm}
\usepackage{amssymb}
\usepackage{colordvi}
\usepackage{color}   
\usepackage{amsbsy}
\usepackage{amsmath}
\usepackage{amsbsy}
\usepackage{latexsym}

\newcommand{\kb}{k_\mathrm{B}}
\newcommand{\br}{\mathbf{r}}
\newcommand{\mism}{d}            
\newcommand{\mismB}{\mathbf{d}}   
\newcommand{\Mism}{\mathbf{D}}    
\newcommand{\bmism}{{\bf d}}  
\newcommand{\avemism}{\bar{d}}   
\newcommand{\xiD}{\xi'_{D}}  
\newcommand{\xiB}{\xi'_{B}}  
\newcommand{\xih}{\xi_{B}}  
\newcommand{\xihL}{\xi_{B,L}}  
\newcommand{\ave}[1]{\langle #1 \rangle}

\begin{document}

\title{Deformation of inherent structures to detect long-range correlations  in supercooled liquids}

\author{Majid Mosayebi$^{1}$, Emanuela Del Gado$^{2}$, Patrick Ilg$^{1}$, and Hans Christian \"Ottinger$^{1}$ }
\affiliation{$^{1}$ ETH Z\"urich, Department of Materials, Polymer
Physics, CH-8093 Z\"urich, Switzerland}
\affiliation{$^{2}$ 
ETH Z\"urich, Department of Civil, Environmental and Geomatic Engineering, Institute for Building Materials, CH-8093 Z\"urich, Switzerland}
\date{\today}

\begin{abstract}
We propose deformations of inherent structures as
a suitable tool for detecting structural changes underlying
the onset of cooperativity in supercooled liquids.
The non-affine displacement (NAD) field resulting from the applied
deformation shows characteristic differences between the high temperature
liquid and supercooled state, that are typically observed in dynamic
quantities. The average magnitude of the NAD is very sensitive to temperature
changes in the supercooled regime and is found to be strongly correlated with the
inherent structure energy.  In addition, the NAD field is characterized by
a correlation length that increases upon lowering the temperature towards
the supercooled regime.

\end{abstract}


\maketitle

\section{Introduction}

Many liquids, when quenched fast enough, enter a supercooled regime that is
signaled by an enormous increase in viscosity eventually leading to glass transition.
Several experimental as well as theoretical studies have shown an accompanying qualitative change of the dynamics: While individual particle dynamics dominate at high temperatures,
particle motions become increasingly more cooperative and heterogeneous as the
temperature is lowered \cite{Wolynes_CRR,Russel_coop,Ediger_coop}.
From the sizes of cooperatively rearranging regions, a growing dynamical length scale can be extracted near the glass transition
\cite{Glotzer_length99,Ludo_Science05,Weeks_correlatedmotion}.

One of the most puzzling features of the supercooled regime and the glass transition
is the apparent lack of structural changes underlying the dramatic slowing down of the dynamics.
Some progress has recently been achieved in linking structural and dynamical changes near the glass transition.
For a two-dimensional model system, a connection between dynamical heterogeneities
and local crystalline order was suggested in \cite{Tanaka_DHandPsi6}.  
Such a connection appears to be absent in a different two-dimensional system,
where instead dynamical heterogeneities seem to be correlated with local,  
short-time dynamics \cite{Harrowell2006} and localized soft modes \cite{WidmerCooperSoftModes}.
A somewhat different path has been followed by several researches,
trying to relate dynamical properties to inherent structures,
which are the local minima of the collective potential energy  
\cite{StillingerWeberIS,StillingerWeberIS2,Debenedetti_review,Heuer_ISreview}.
It has been shown that inherent structures govern mechanical properties
of amorphous solids \cite{Malandro,Lemaitre06,Fabien_prl}.
For supercooled liquids, more and more evidence has been collected
that qualitative changes in the dynamical behavior are accompanied by changes
in the inherent structures \cite{SastryIS,SastryIS_fragile,Debenedetti_review}.
The local diffusivity, for example, was suggested to be related to the basin depth
of local inherent structures \cite{LaNave_DiffandIS}.  
Similarly, unstable modes of saddle points in the energy surface were found to be
of major importance, not only for the mobilities of particle clusters
\cite{Pastore_localsaddle} but for the glass transition in general
\cite{Parisi_geometricGlass}.
Collective rearrangements in the inherent structures corresponding to the
dynamics of supercooled liquids are reported in \cite{Bailey_avalancheinIS}.
Specially designed simulations probing point-to-set correlations
that are inspired by random first order theory have revealed evidence
for growing amorphous order in the supercooled regime \cite{Biroli_overlapRFOT}.
In some very recent works, external forcing is applied to (some parts of) the system
in order to extract characteristic sizes of cooperative regions using
mode-coupling \cite{inhomMCT} or density functional theory \cite{Yoshidome}.
Under shear flow, it was found numerically that cooperative, mobile regions can form
anisotropic bands \cite{FuruTanaka_anicooperative09}.

We have found striking similarities in the onset of cooperative behavior
in dynamics and in the non-affine part of the inherent structure response
to external deformations \cite{ema_mismatch}.
That approach was motivated by a recent theory \cite{hco_glass}
based on a general framework of non-equilibrium thermodynamics.
This thermodynamic treatment requires information about the system's response
to an applied, static deformation and  
suggests that the reversible part of glassy dynamics changes considerably
when approaching the glass transition.
Above the glass transition, the particles can follow an imposed deformation
more or less freely, whereas closer to the glass transition
the particle movement becomes a hopping-like transition between
different basins of attraction of the underlying inherent structures
\cite{StillingerWeberIS,schroder_ISdynmamics}.
By imposing static deformations of inherent structure configurations,
we observed indeed a profound difference in the NADs when
approaching the glass transition.
From a systematic finite-size scaling analysis, we found that the
NAD field is characterized by a static correlation length,
that is growing as in usual critical phenomena \cite{Majid_criticallength, Majid_thesis}.
This length detects growing structural correlations underlying the
growing dynamical length scale obtained from particle dynamics
\cite{Glotzer_length99,Ludo_Science05,Sastry_growinglength}.

Here we present an extensive study of the NAD field introduced in
\cite{ema_mismatch,Majid_criticallength}, that further demonstrates how  
this quantity can effectively detect the onset of the cooperative dynamics and shows a great potential 
to probe the underlying structural correlations. The distribution of lengths changes from
exponential to power law upon entering the supercooled regime, and we can rationalize
such change in terms of a crossover from a viscous liquid to a regime dominated by
elastic effects. The mean displacement length depends exponentially on the inherent
structure energy, as we also discuss using a simple toy model, and confirms the
existence of two well-distinguished regimes as a function of the temperature.
We analyze different measures of correlations in the direction of the NAD field and discuss
their analogies with observations in the dynamics \cite{Weeks_correlatedmotion}.
Finally we use a coarse-graining procedure to extract the characteristic size of
correlated regions observed in the snapshots: we discuss different definition of this
length-scale and we perform a finite-size scaling analysis over different model systems,
confirming the critical-like behavior at low temperatures.

The paper is organized as follows. The model glass formers used and the numerical simulations 
are described in Sect.~\ref{sec:model}. 
In Sect.~\ref{sec:method}, we briefly review the method proposed in \cite{ema_mismatch} 
to extract displacements of inherent structures.
Our numerical results for the NAD are presented in 
Sects.~\ref{sec:results1} and \ref{sec:results2}. 
Correlations between these displacements are analyzed in Sect.~\ref{sec:results2}. 
We focus in Sec.~\ref{sec:results1} on characterizing the NAD lengths and we introduce 
a simple model to rationalize the results. In Section~\ref{sec:coarsegr} we extensively describe  
the coarse graining procedure for the analysis of the NADs introduced in 
Ref.~\cite{Majid_criticallength} and apply it to the different model systems to extract 
the temperature and system size dependence of the correlation length.
Sect.~\ref{sec:concl} contains further discussion and conclusions.

\section{Model description} \label{sec:model}
The studies presented in this paper are based on computer simulations of
two different, well-established models for structural glasses  
(see e.g.~\cite{Andersen_review05}).

The first model is the three-dimensional,
binary mixture of Lennard-Jones (BMLJ) particles introduced by Kob and Andersen
\cite{KobAndersen_1995a}. Both particles have unit mass $m$ and interact via a Lennard-Jones potential,  
$\Phi_{ab}(r) = 4\epsilon_{ab}[(\sigma_{ab}/r)^{12}-(\sigma_{ab}/r)^6 ]$, 
where the diameters and interaction energies
are given by
$\sigma_{AA}=\sigma_0$, $\sigma_{AB}=0.8\sigma_0$, $\sigma_{BB}=0.88\sigma_0$,
$\epsilon_{AA}=\epsilon$, $\epsilon_{AB}=1.5\epsilon$, and
$\epsilon_{BB}=0.5\epsilon$. We have chosen $\sigma_0$ as the unit of length and $\epsilon$ as the unit of energy.
The potential is truncated and shifted to ensure $\Phi_{ab}(r_\mathrm{cut})=0$,
where the cut-off distance is chosen as $r_\mathrm{cut}=2.5\sigma_{ab}$.

The second model considered in this study
is a 50:50 binary mixture of soft spheres (BMSS) in three dimensions \cite{Biroli_overlapRFOT}.
Both particle types have unit mass and the interaction potential is given by
$\Phi_{ab}(r)=\epsilon(\sigma_{ab}/r)^{12}$,
$\sigma_{ab}=\sigma_a+\sigma_b$,  
$a,b=\{A,B\}$,
where the sizes of particles $\sigma_{ab}$ are fixed by setting $\sigma_A/\sigma_B=1.2$ and the effective diameter to one; that is
$(2\sigma_A)^3+(2\sigma_B)^3+2(\sigma_A+\sigma_B)^3=4\sigma_0^3$. $\sigma_0$ 
is the unit of length and density is chosen to be $\rho=N/V=\sigma_0^{-3}$.  
A smooth cut-off is used by setting the potential to
$\Phi_{ab}(r)=B_{ab}(a-r_\mathrm{cut})^3+C_{ab}$ for $a>r>r_\mathrm{cut}=\sqrt3$
and
$\Phi_{ab}(r)=C_{ab}$ for $r>a$.
The values of $B_{ab}$, $C_{ab}$ and $a$ are fixed by imposing the continuity up to second derivative for $\Phi_{ab}(r)$.
$$B_{ab}=\frac{169 \epsilon}{r_\mathrm{cut}^3}(\frac{\sigma_{ab}}{r_\mathrm{cut}})^{12},$$
$$C_{ab}=\frac{5 \epsilon}{13}(\frac{\sigma_{ab}}{r_\mathrm{cut}})^{12},$$
$$a=(15/13)r_\mathrm{cut}.$$
The potential is then shifted to ensure that $\Phi_{ab}(a)=0$.

\begin{center}
\begin{table}
\begin{tabular}{c|c|c|c|c}
abbr. &  potential $\Phi_{ab}$ & A:B & $T_{\rm K}$ & $T_{\rm MCT}$ \\
\hline
BMLJ & $4\epsilon_{ab}[(\sigma_{ab}/r)^{12}-(\sigma_{ab}/r)^6 ]$ & 80:20 & $\approx 0.30$ & $0.435$ \\
BMSS & $\epsilon (\sigma_{ab}/r)^{12}$                   & 50:50 & $\approx 0.11$ & $\approx 0.226$  
\end{tabular}\\
\caption{Some parameter values for the two models used. $T_{\rm K}$ denotes the (extrapolated) Kauzmann temperature 
and $T_{\rm MCT}$ is the mode-coupling temperature. 
The mode coupling temperature is estimated in Ref.~\cite{KobAndersen_1995a} and \cite{Hansen_BMSS}
for the BMLJ and BMSS models, respectively.
For the BMLJ model, $T_{\rm K}$ has been
estimated numerically as $T_{\rm K}\approx 0.30$ \cite{Sciortino_IS99,Coluzzi00}.
There have been several theoretical and numerical estimates for $T_{\rm K}$ in the BMSS model; 
$0.11 \lesssim T_{\rm K} \lesssim 0.14$ have been reported in Ref.~\cite{Coluzzi99,yoshino}. 
Here we take the numerical estimate given in Ref.~\cite{Coluzzi99}. 
The onset temperature $T_a$ of the non-Arrhenius behavior of transport properties is estimated to be $T_a\approx 1.0$
for the BMLJ model. $T_a$ depends somewhat on the quantities investigated and is to be taken as a rough
estimate. A careful investigation of $T_a$ for the BMLJ model is presented in \cite{Sastry_TainKA}. }
\label{table:models}
\end{table}
\end{center}
Main parameter values for the two models are shown in Table \ref{table:models}.
For ease of notation, we use the same symbols for original
and reduced quantities.

For both model systems studied here, we have carefully prepared 10-100 independent samples
for each temperature by molecular dynamics simulations
starting from statistically independent, random initial configurations.
Periodic boundary conditions were used in all cases.
We have studied systems with number of particles $N$ varying from $2000$ up to $64000$, 
but most of the data here refer to systems with $N=8000$. The  simulations were performed with the molecular
dynamics simulation package LAMMPS \cite{lammps}.
The initial particle configurations were equilibrated at several decreasing
values of temperature in the range $0.20\leq T\leq 1.0$ (BMSS), and
$0.42\leq T\leq 3.5$ (BMLJ). Comparing these temperature intervals with the mode-coupling temperature $T_{\rm MCT}$ given in
table \ref{table:models}, our simulations cover the high temperature regime down to the supercooled state and 
extend below the mode-coupling temperature.
For all systems, slowly cooling the configurations towards low temperatures
was achieved by coupling the system to Nos\'e-Hoover thermostat
with prescribed, slowly decreasing temperature protocol.
When the target temperature was reached, the temperature of the thermostat was
held constant and the system was equilibrated at this temperature 
still in contact with the thermostat. We verified that no significant drift in the internal energy nor 
in the kinetic temperature could be observed after the thermostat was switched off and the system was 
further evolved, now in the microcanonical ensemble.
To test the equilibration of samples, we also compared the kinetic ($T_\mathrm{kin}$) with 
the configurational temperature ($T_\mathrm{conf}$) \cite{Evans_Tconf}.
The kinetic temperature is defined by
$N_f\kb T_\mathrm{kin}=\sum_{j=1}^Nm{\bf v}_j^2$,
where $N_f$ denotes the number of degrees of freedom.
While the kinetic temperature is entirely determined by the particle velocities,
the configurational temperature $T_\mathrm{conf}$ depends on the positions of the
particles via $\kb T_\mathrm{conf}=\ave{|\sum_i {\bf F}_i|^2}/\ave{-\sum_j \nabla_j\cdot{\bf F}_j}$.
Here, ${\bf F}_j$ denotes the total force on particle $j$ and
$\nabla_j=\partial/\partial {\bf r}_j$. We verified that kinetic and configurational temperature agree
for our equilibrated samples within numerical uncertainties.
Finally, we calculated two-time correlation functions and verified that no significant 
aging was observed in the equilibrated samples for a waiting time of the order of 
$40 \tau_\alpha$, where 
$\tau_\alpha$ is the structural relaxation time.

\section{Method for extracting NAD} \label{sec:method}
To extract non-affine particle displacements we proceeded as proposed in
our earlier work \cite{ema_mismatch}. In order to make the paper self-contained
and settle the notation, we briefly review this method here.

\subsection{Affine deformations}
\begin{figure}[h]
\vskip 0.5cm
\includegraphics[width=1.00\linewidth]{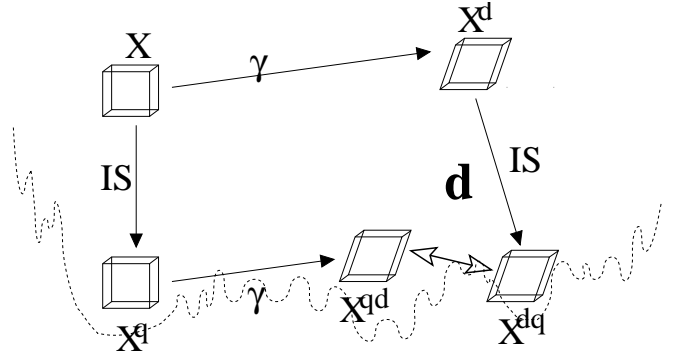}\\
  \caption{Schematic plot of the preparation of the configurations  
$X^\mathrm{qd}$ and $X^\mathrm{dq}$ and their NAD $d$.}
  \label{quenchdeform.fig}
\vskip 0.5cm
\end{figure}

We apply static, affine shear deformations to the particle configuration by mapping the particle
positions $\br_i \to \br_i^\mathrm{d}$,  
with $\br_i^\mathrm{d} = \br_i + \gamma y_i {\bf e}_x$,
where $\gamma$ denotes the deformation amplitude.
Since we aim at the non-affine part of the inherent structure
deformations \cite{hco_glass}, we suggested in \cite{ema_mismatch}
the following procedure, summarized schematically in Fig.~\ref{quenchdeform.fig}:  
start with configuration $X=\{\br_i\}$ at a given temperature
$T$. Prepare one configuration $X^\mathrm{dq}$ by first
applying the static deformation $\br_i \to \br_i^\mathrm{d}$
and subsequently finding the inherent structure corresponding
to this deformed configuration.
The other configuration $X^\mathrm{qd}$ is prepared by
first finding the inherent structure corresponding to the
initial configuration $X$ and after that subjecting the
inherent structure configuration to the same deformation.
Almost all subsequent analysis is based on the comparison between
the configurations $X^\mathrm{dq}$ and $X^\mathrm{qd}$,
which we denote as non-affine displacement (NAD), 
$\bmism_j\equiv \br_j^\mathrm{dq}-\br_j^\mathrm{qd}$.
Thereby, we focus on the dependence of the NAD on  
the temperature $T$ of the initial configuration and the
amplitude $\gamma$ of the applied deformation.
We ensure that the total displacement vanishes, $\sum_j\bmism_j={\bf 0}$, 
since a rigid translation can always be added and does not contribute to the NAD.

\subsection{Inherent structure generation}
From the equilibrated samples, 
we generate inherent structure (IS) configurations by minimizing the
potential energy via the conjugate gradient method
\cite{Stillinger_return}.
The minimization is stopped when the potential energy change is less than
a tolerance value $10^{-7}\epsilon$. We verified that the results are
insensitive to a further decrease of the tolerance level and that the mean inherent structure
energies for different temperatures agree well with published data
\cite{dePablo_DOSMCglass,SastryIS,Cammarota_spinodal09}.
For generating the inherent structure of deformed configurations, $X^\mathrm{dq}$,
the minimization is performed in a deformed simulation box or,
equivalently, using Lees-Edwards boundary conditions.

\subsection{Inherent structure properties}
\begin{figure}[h]
\vskip 0.5cm
\includegraphics[width=1.00\linewidth]{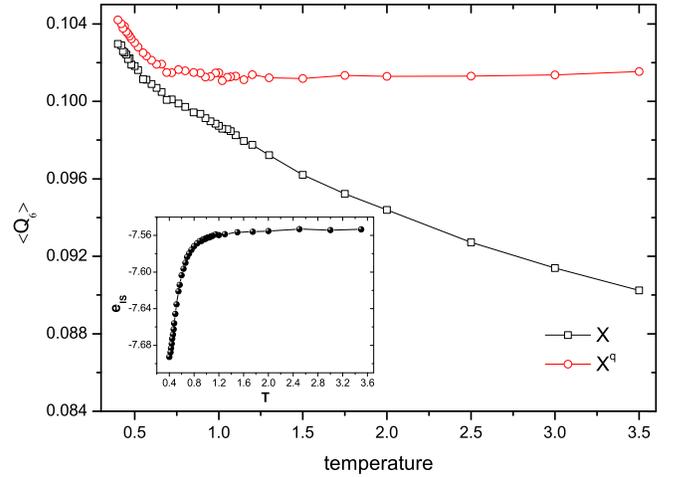}\\
  \caption{Main frame: Average bond orientational order measured in the starting configuration $X$ 
  and its inherent sructure $X^\mathrm{q}$ as a function of temperature for BMLJ model. 
  Inset: Average energy of the inherent structure configuration as a function of temperature $T$ for BMLJ model. 
  Error bars are smaller than the size of the symbols.}
  \label{Q6.fig}
\vskip 0.5cm
\end{figure}

Inherent structures can be characterized by their mean energy $e_{\rm IS}$. 
In the inset of Fig.~\ref{Q6.fig} we plot $e_{\rm IS}$ of the inherent structure $X^{\rm q}$ 
as a function of $T$. 
Starting from a high temperature plateau, $e_{\rm IS}$ decreases upon cooling in 
the so-called landscape dominated regime \cite{Debenedetti_review}.
In addition, we have performed various types of structural analysis on the IS of different 
model systems, using for example pair correlation functions, coordination numbers or bond order parameters, 
to characterize the differences between the particle configurations in the initial state and in the IS.
As an example, in Fig.~\ref{Q6.fig} we plot the averaged bond orientational order parameter 
 $\ave{Q_6}$ for the model BMLJ.  $\ave{Q_6}$ is calculated by averaging, for each temperature, 
over $100$ independent samples of $N=8000$ particles, the $l$-th order 
bond orientational order parameter \cite{Steinhardt} with $l=6$. 
For particle $j$, this quantity is calculated as
\begin{equation}
Q_{l}^{j} = \sqrt{\frac{4\pi}{2l+1}} \left( \sum_{m=-l}^{l} | Q_{lm}^{j} | ^{2} \right)^{1/2}
\end{equation} 
where $Q_{lm}^{j}$ is the locally averaged bond orientational order parameter of order $l$ and 
degree $m$ as is defined in \cite{Steinhardt,Tanaka_critical}.
 In the main frame of  Fig. \ref{Q6.fig}, $\langle Q_{6}\rangle$ is plotted as a function of temperature 
for the starting configuration $X$ and its inherent structure $X^{\rm q}$. 
We observe that, in the range of temperatures where $e_{\rm IS}$ 
displays a strong dependence on $T$, the $6$th order averaged bond orientational order clearly increases
 for both $X$ and $X^{\rm q}$. 
Also, as expected, the value obtained for $X^{\rm q}$ is 
larger. In spite of this, the values of $\langle Q_{6} \rangle$ are consistently much lower than those 
expected for various crystal lattices (e.g.~$Q_6\approx 0.66$ for perfect icosahedral ordering),
demonstrating that there 
is no significant crystallinity present
in the $X$ and $X^{\rm q}$ configurations \cite{Aste}.

\section{Characterizing the NAD field} \label{sec:results1}
We have applied the procedure described in Sec.~\ref{sec:method} to well-equilibrated
configurations of the two model systems.
Various temperatures of the initial configurations as well as different deformations
are considered. Results are obtained as averages over independently generated configurations and error bars 
are calculated from the standard deviation.

\subsection{Distribution of NAD vectors}
\begin{figure}[h]
\vskip 0.5cm
\includegraphics[width=1.00\linewidth]{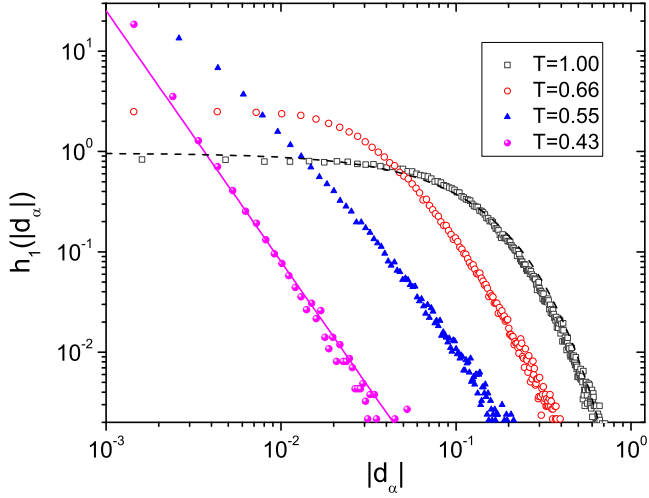}\\
  \caption{Distribution of mismatch components $h_1$ (defined in the text) in the BMLJ model
  for homogeneous deformation of strength $\gamma=10^{-4}$.
  Temperature is increasing as
  $T=0.43, 0.55, 0.66, 1.0$ from left to right at the bottom of the figure.
  At high temperatures the distributions are exponential (dashed line shows a fit 
  $\exp{[-|d_\alpha|/0.1]}$ to the data at $T=1.0$),
  but start to develop a power-law tail at low temperatures.
  The solid line shows a fit to a power-law with an exponent $-5/2$.
}
  \label{mismatchdistrib.fig}
\vskip 0.5cm
\end{figure}

We start by characterizing the distribution of the NAD vectors
$h(\bmism)$.
For the small deformations $10^{-5}\lesssim\gamma\lesssim 10^{-2}$ we are interested in,
the distribution $h(\bmism)$ is found to be isotropic for homogeneous deformations 
and covers a range of displacement magnitudes.
We therefore conclude, that our regime of deformations is strong enough to
allow the escape from local minima.  On the other hand, the deformations are small enough 
such that no obvious trace of the imposed shear geometry is seen in $h(\bmism)$.
This can be seen from Fig.~\ref{mismatchdistrib.fig}, which shows
the distribution of the Cartesian NAD components $h_1(|\mism_\alpha|)$, $\alpha=1,2,3$.  
The distributions $h_1$ show a maximum at $\mism_\alpha=0$.
At high temperatures, $h_1(|\mism_\alpha|)$ decays exponentially, 
$h_1(|\mism_\alpha|)\propto \exp{[-|\mism_\alpha|/\avemism_{\rm c}]}$,
with fit parameter $\avemism_{\rm c}\approx 0.1$ of the order of the mean NAD length at this temperature.
This behavior is reminiscent of
the exponential distribution observed in Ref.~\cite{Bailey_avalancheinIS},
also with $\avemism_{\rm c}\approx 0.1$, for displacements between inherent structures corresponding
to the system's dynamics.

\begin{figure}[h]
  \vskip 0.5cm
\includegraphics[width=0.5\textwidth]{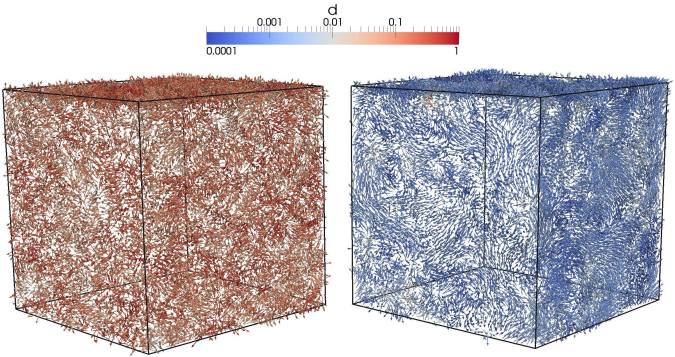}\\
  \caption{NAD field for for one selected configuration at high ($T=1.0$ left) and low ($T=0.42$ right)
  temperature. Homogeneous deformation of strength $\gamma=10^{-4}$ was applied to BMLJ system. For clarity just particles on visible surfaces are shown.}
  \label{snap_mismatch.fig}
\vskip 0.5cm
\end{figure}

This finding indicates again that the non-affine inherent structure deformations
bear remarkable similarities to dynamical properties.
At low temperatures, the distribution $h_1$ is much narrower, in agreement with
the impression from the snapshots shown in Fig.~\ref{snap_mismatch.fig}.
More quantitatively, one finds that the distribution at low temperatures is no
longer decaying exponentially but seems to exhibit power-law tails
$h_1(|\mism_\alpha|)\propto |\mism_\alpha|^{-\nu}$ with an exponent approaching
$\nu\approx 2.5$ at the lowest temperatures investigated,
see the solid line in Fig.~\ref{mismatchdistrib.fig}.
It is interesting to note, that the power-law distribution
$h_1(|\mism_\alpha|)\propto |\mism_\alpha|^{-5/2}$
is predicted for the induced radial displacements in an elastic continuum around an  
expanded spherical shell \cite{Dyre99,Dyre99_II}.
If we can identify the latter with localized rearrangements,
the qualitative change of the NAD distribution from exponential to a power-law
at lower temperatures apparently indicates the crossover from a viscous liquid to
a regime with more pronounced elastic effects. 
From the goodness-of-fit, we determine the crossover temperature where the power-law 
and exponential distributions fit equally well to the data. We find the crossover temperature to be 
located in the interval $T\in(0.65, 0.70)$ for the BMLJ model (data not shown).

\subsection{Mean Mismatch length} \label{sec:meanlength}
\begin{figure}[h]
\vskip 0.5cm
\includegraphics[width=1.00\linewidth]{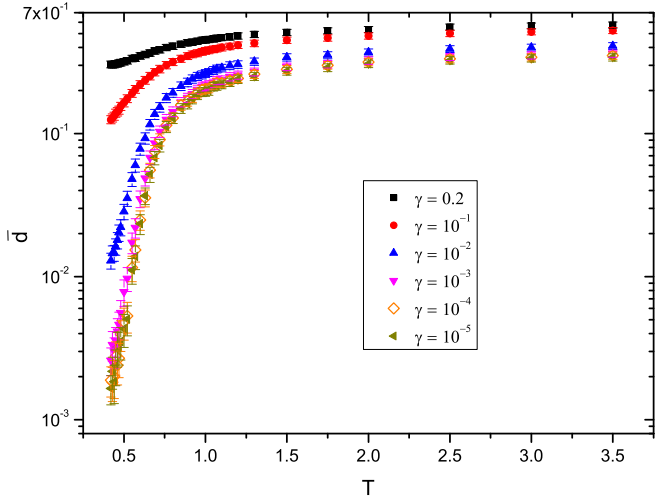}\\
  \caption{The mean length of the mismatch vectors $\avemism$ defined in the text
as a function of temperature for the BMLJ model and for different deformation magnitude $\gamma$. 
}
  \label{dlength_T.fig}
\vskip 0.5cm
\end{figure}

A first characterization of the NAD field $\{\bmism_j\}$
is provided by their mean length,
\begin{equation} \label{dlength}
\avemism(T,\gamma) = \ave{N^{-1}\sum_j\bmism_j^2}^{1/2},
\end{equation}
where the ensemble average is performed over the independently 
generated samples.
Figure \ref{dlength_T.fig} shows $\avemism$ as a function of temperature $T$
for homogeneous deformations of different strengths $\gamma$ for the BMLJ model.
Depending on the value of $\gamma$, we distinguish three different regimes.
For small values of $\gamma$ ($\gamma\lesssim 10^{-4}$), we find
that $\avemism$ is essentially independent of $\gamma$. 
This is an interesting observation as it suggests
that we have reached a regime of weak perturbations,
where details of the local deformation are less important for $\avemism$.
Moreover, $\avemism$ decreases drastically, by roughly two orders of magnitude, as
the temperature is lowered from $T\gtrsim 1$ to $T\approx 0.45$.
For larger deformations $10^{-3}\lesssim \gamma \lesssim 10^{-1}$, the decrease
of $\avemism$ gets less and less pronounced, until
$\avemism$ becomes essentially independent of temperature for deformations
$\gamma\gtrsim 0.1$.
The particular form of the drastic decrease in $\avemism(T)$ depends on the model
system.
Qualitatively, however, the observations are the same also for the BMSS model.

The decrease of $\avemism$ with temperature for small deformations shown in
Fig.~\ref{dlength_T.fig} is quite similar to the decrease of the inherent structure
energy $e_\mathrm{IS}$ 
(see \cite{Heuer_ISreview} and the inset of Fig.~\ref{dlength_T.fig}).
Therefore, we parametrically plot in Fig.~\ref{dlength_eIS.fig} the average NAD length
$\avemism$ versus $e_\mathrm{IS}$ for the different temperatures investigated. 
Our result indicate that $\avemism$ 
is indeed strongly coupled to the inherent structure energy.
From Fig.~\ref{dlength_eIS.fig} we observe that
$\avemism$ increases exponentially with the inherent structure energy.
More quantitatively, we observe a crossover from
$\avemism\simeq e^{-\beta_1|e_\mathrm{IS}|}$, $\beta_1=22.5$ at low temperatures to
$\avemism\simeq e^{-\beta_2|e_\mathrm{IS}|}$, $\beta_2=61.9$ at higher temperatures. 
Since the mean inherent structure energies vary inversely proportional with
temperature in the landscape-influenced regime \cite{Heuer_ISreview}, we find that
$\avemism\simeq e^{-A/T}$ with a different constant $A$ in the two regimes.
The crossover from $\beta_1$ to $\beta_2$
takes place at $e_\mathrm{IS}\approx -7.63$, which corresponds
to a temperature $T\approx 0.6$, near to the inflection point of
$e_\mathrm{IS}(T)$, see the inset of Fig.\ \ref{dlength_T.fig}.

\subsection{Simple toy model for inherent structure NAD} \label{sec:toymodel}

Several features of the NAD field can be rationalized using a simple toy model.
Let $X$ denote again the state of our system and $X^\mathrm{q}$ its closest minimum
(inherent structure). 
For simplicity, however, the toy model describes these states by a single scalar variable.
The imposed deformation of strength $\gamma$ deforms these configurations
to $X^\mathrm{d}=X+g$ and $X^\mathrm{qd}=X^\mathrm{q}+g$, respectively, where $g$ denotes
an average amount of displacement.
The inherent structure of the deformed system is
$X^\mathrm{dq}=X^\mathrm{q}+n\Gamma$, where $\Gamma$ denotes a typical size of
an inherent structure basin. The integer $n$ takes values $n=0,1,\dots$,
with $n=0$ corresponds to staying in the same minimum, $n=1$ to the nearest
minimum, etc.
The NAD $\mism$ is therefore given by $\mism=n\Gamma-g$ and its mean squared
average by
\begin{equation} \label{mismatch_toymodel}
\avemism = [\sum_n p_n (n\Gamma - g)^2]^{1/2}
\end{equation}
In Eq.~(\ref{mismatch_toymodel}), we have introduced the
probabilities $p_n$ of finding $X^\mathrm{dq}$ in the $n$-th nearest minimum.
In the limit of $p_n\to \delta_{n,0}$,
we find that $\avemism=g$, i.e.~the NAD length is given by the
imposed deformation.
This is the case for extremely low temperatures and/or vanishing deformations, 
not considered in the present study.
For very large deformations compared to the basin size $g\gg \Gamma$,
we expect equal probability of entering a basin $n$ in the neighborhood
$\Delta m$ around $m=g/\Gamma$, where $m\gg 1$. Therefore,  
$p_n\approx 1/2\Delta m$ for $|n-m|<\Delta m$ and zero else.
Inserting this expression into Eq.~(\ref{mismatch_toymodel}) we find
$\avemism\approx g \Delta m/m\to \Delta m \Gamma$.
Hence, in the case of very large deformations, the average NAD is expected to 
approach a limiting value. 
The simulation data in Fig.~\ref{dlength_T.fig} seem to indicate such a trend.
However the mean length remains weakly sensitive to temperature variations at the 
largest deformation investigated.

\begin{figure}[h]
\vskip 0.5cm
\includegraphics[width=1.00\linewidth]{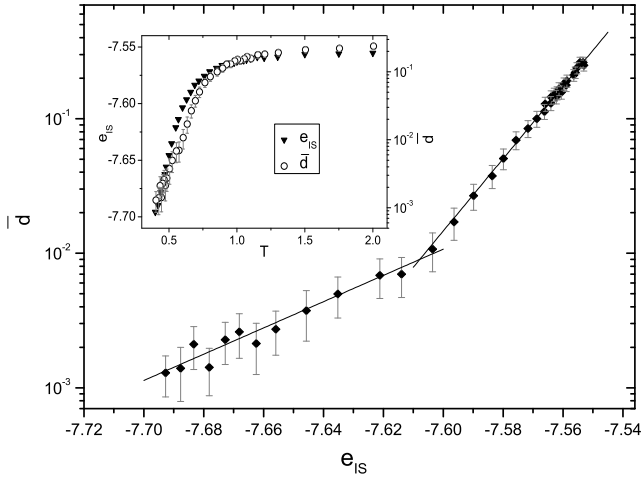}\\
  \caption{Main frame: The mean length of NAD vectors as a function of inherent structure energy $e_\mathrm{IS}$ for the BMLJ model.
           Note the crossover from high temperature to low temperature regime which happens slightly below $T_a$.
           Inset: temperature dependence of $\avemism$ and $e_\mathrm{IS}$. }
  \label{dlength_eIS.fig}
\vskip 0.5cm
\end{figure}

Finally, we consider the case of rather small deformations,
where one can approximately set $p_n=0$ for $n>1$.
The important quantity is then $p_1$, the probability of leaving the basin
at the imposed deformation $\gamma$.
Accounting in a rough way for the equilibrium distribution within the basin,
we assume $p_1=\gamma e^{-\beta \epsilon}$, where $\epsilon$ denotes a typical  
energy barrier and $\beta=(\kb T)^{-1}$.
The values of $\gamma$ are restricted to $\gamma\leq 1$ in order to
ensure proper probabilities $p_1\geq 0$ for all temperatures.
With these assumptions, the mean NAD becomes
\begin{equation} \label{mismatch2_toymodel}
\avemism=\Gamma[\gamma(1-2\gamma)e^{-\beta\epsilon}+\gamma^2]^{1/2},
\end{equation}
where we have assumed $g=\gamma\Gamma$, i.e.~the important length for the imposed displacement
is the typical basin size.
Equation (\ref{mismatch2_toymodel}) predicts a number of features that can be tested  
by simulations.
First, for high temperatures, $\avemism$ reaches a limiting value
$\avemism\to \Gamma\sqrt{\gamma(1-\gamma)}$
independent of temperature, in agreement with simulation results \cite{Majid_thesis}.
This limiting value decreases as $\gamma$ increases towards its maximum value.
Second, for small deformations $\gamma\ll 1$, Eq.~(\ref{mismatch2_toymodel}) simplifies to
$\avemism=\Gamma\sqrt{\gamma}\exp{[-\beta\epsilon/2]}$ and predicts a square-root
increase of $\avemism$ with the strength $\gamma$ of imposed deformation.
In the simulations for the BMLJ model, $\avemism$ indeed increases with $\gamma$,
the exponent, however, changes with temperature from $0.2$ at $T=1$ to $1.0$ at $T=0.42$ \cite{Majid_thesis}.
Third, when the temperature is decreased, Eq.~(\ref{mismatch2_toymodel}) predicts a decrease
of $\avemism$ according to the Boltzmann factor. This is in agreement with the results 
discussed in Sec.~\ref{sec:meanlength} (see also Fig.\ref{dlength_eIS.fig}).
Finally, Eq.~(\ref{mismatch2_toymodel}) shows an exponential dependence of
the average NAD on the barrier height $\epsilon$.
If we can associate the inherent structure energy with an effective barrier height,  
this prediction is consistent with our numerical results
(see Fig.~\ref{dlength_eIS.fig}).
More quantitatively, we assume that the effective barriers are given by
$\epsilon(T)=-2\beta_1\kb T(e_{\rm IS}(T)-e_{\rm IS}^\infty)$, where
$e_{\rm IS}^\infty$ is the high-temperature plateau of the inherent structure energy
and the coefficient $\beta_1$ describes the dependence of $\avemism$ on $e_{\rm IS}$ in the low temperature regime, 
as introduced at the end of Section \ref{sec:meanlength}.
A linear relation between mean inherent structure energies
and barrier heights was indeed found numerically for low temperatures
in the BMLJ model \cite{DoliwaHeuer} and
similarly in a binary soft sphere model \cite{Parisi_geometricGlass}.
For low temperatures, the mean inherent structure energy decreases as
$e_{\rm IS}=e_0 - \sigma^2/\kb T$, 
where $\sigma^2$ can be interpreted as the variance of the inherent structure energy distribution in 
the Gaussian energy landscape model 
\cite{Heuer_ISreview}.
In fact, we find in this regime that
the height of effective barriers decrease linearly with temperature,
$\epsilon(T)=\epsilon_0-\alpha\kb T$ for $\kb T < \epsilon_0/\alpha$,
where $\epsilon_0=2\beta_1\sigma^2$ and $\alpha=2\beta_1(e_0-e_{\rm IS}^\infty)$.
For the BMLJ model, we have
$e_{\rm IS}^\infty\approx -7.55, e_0\approx -7.39, \sigma^2\approx 0.12, \beta_1\approx 22.5$,
resulting in
$\epsilon_0\approx 5.4$ and $\alpha\approx 7.6$.
The value of the extrapolated barrier height
$\epsilon_0$ is close to the one
obtained from the mean waiting time for low-lying IS \cite{DoliwaHeuer}.

Within the toy model Eq.~(\ref{mismatch2_toymodel}), the crossover from low to high
temperature behavior observed in
Fig.~\ref{dlength_eIS.fig}
could be interpreted as different regimes in the potential energy
landscape, with different relations between minima (inherent structures) and barrier
heights.
Interestingly, numerical simulations of randomly perturbed inherent
structures for BMLJ model indicated that the number of saddles vanishes exponentially
below $T\lesssim 0.6$ \cite{fabricius_distanceIS}, close to the temperature corresponding to the inflection point 
shown in Fig. \ref{dlength_eIS.fig} as evaluated at the end of Section \ref{sec:meanlength}.

\section{Correlations in the NAD field} \label{sec:results2}

Not only the length of the NAD vectors but also their spatial correlation
change considerably between the liquid and supercooled regime.
Spatial correlations in the NAD field can be clearly seen in the snapshots in
Fig.~\ref{snap_mismatch.fig} and were also reported in our earlier
study \cite{ema_mismatch}. In the following we thoroughly analyze them and 
use them to extract correlation lengths.

\subsection{Correlated mobility and directions}
We quantify correlations in the NAD field over a given distance $r$ and
for deformation amplitude $\gamma$ by
\begin{subequations} \label{WeeksCorrel}
\begin{eqnarray}
C_{\delta}(r,\gamma) & = & \frac{\langle \sum_{i,j} \delta \mism_{i}\, \delta \mism_{j} \delta(r-r_{ij})\rangle}{\langle (\delta \mism)^{2} \rangle}, \\
C_{\|}(r,\gamma) & = & \langle \sum_{i,j}{{\mism}}_{i}^\| \, {{\mism}}_{j}^\| \delta(r-r_{ij})\rangle, \\
C_{\perp}(r,\gamma) & = & \langle \sum_{i,j}{{\bmism}}_{i}^\perp \cdot {{\bmism}}_{j}^\perp \delta(r-r_{ij})\rangle,
\end{eqnarray}
\end{subequations}
where $r_{ij}$ denotes the distance between
particles $i$ and $j$ and the average in the denominator is performed over all particles. 
The normalization is chosen such that $C_\delta=1$, $C_\|=1/3$ and $C_\perp=2/3$ 
for perfectly correlated particles
and  $C_{a}=0$ for uncorrelated displacements, $a=\{\delta,\|,\perp\}$.
The same correlation functions were used in Ref.~\cite{Weeks_correlatedmotion}
in order to study correlated displacements in hard sphere colloids.
The first function, $C_{\delta}$ measures correlations in the mobility, i.e.~the magnitude
of the particle displacements (with the average subtracted),
$\delta \mism_{j} = |\bmism_{j}| - \langle|\bmism|\rangle$,
whereas $C_{\|,\perp}$ is sensitive to correlated directions.
The latter distinguish longitudinal and transverse correlations of directions,
with $\mism_{i}^\|\equiv\hat{\bmism}_{i}\cdot\hat{\br}_{ij}$ and
${\bmism}_{i}^\perp=\hat{\bmism}_{i} - \hat{\mism}_{i}^\|\hat{\br}_{ij}$,
with $\hat{\br}_{ij}$ the unit vector connecting particles $i$ and $j$ and $\hat{\bmism}_{i}=\bmism_i/
|\bmism_i|$.

\begin{figure}[h]
\vskip 0.5cm
\includegraphics[width=1.0\linewidth]{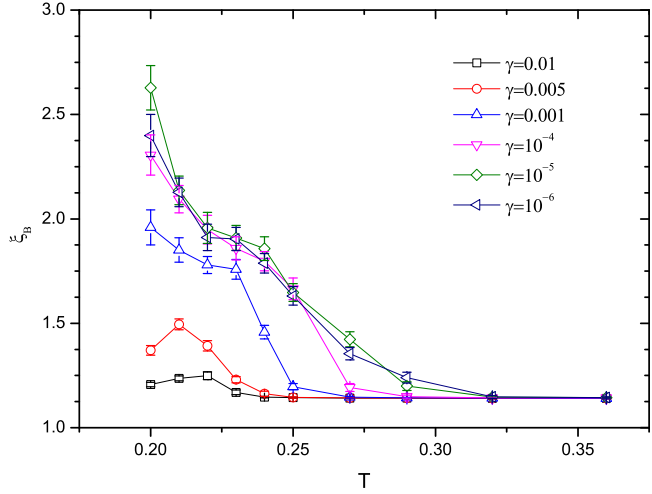}\\
  \caption{The length scale extracted from peak-location of histograms of coarse-grained NAD orientations
           as function of temperature $T$ for different strength of shear deformation $\gamma$ for BMSS model with $N=16000$. At low $\gamma$ the length becomes
           independent of deformation magnitude and hence an intrinsic property of the system.}
  \label{xi_gamma.fig}
\vskip 0.5cm
\end{figure}

Fixing the particle separation around the first neighbor shell and varying the deformation amplitude
$\gamma$, we find that all three correlation functions (\ref{WeeksCorrel}) are rather insensitive
to $\gamma$ and start to decay only for $\gamma\gtrsim 0.1$ (not shown).
This value is consistent with the treshold value determined in Ref.~\cite{ashwin} for shear-induced 
transitions between different inherent structures. 
Since we observe correlated rearrangements at smaller $\gamma$, 
we would rather conclude that such large
deformations induce transition between uncorrelated metabasins 
(see also Fig.~\ref{xi_gamma.fig}).
The behavior of $C_{\|}$ and $C_{\perp}$, as well as the relation
$C_{\delta}>C_{\|}+C_{\perp}$ found, are quite similar to the one observed in
Ref.~\cite{Weeks_correlatedmotion} if our deformation $\gamma$ is
identified with a properly scaled time (for not too short times).

\begin{figure}[h]
\vskip 0.5cm
\includegraphics[width=1.00\linewidth]{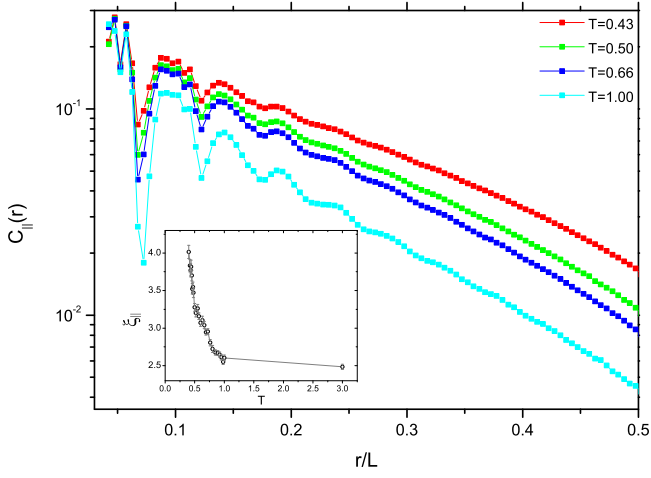}\\
  \caption{Correlation of longitudinal mismatch directions $C_{\|}$ as a function of particle separation
  $r$ defined in Eq.~(\ref{WeeksCorrel}).
  Homogeneous shear deformation with $\gamma=10^{-4}$ was applied to BMLJ model with $N=8000$ particles.
  Inset shows the length scale obtained by fitting an exponentially decaying function to data.}
  \label{Cpara.fig}
\vskip 0.5cm
\end{figure}

Next, we fix $\gamma$ to a typical small value ($\gamma=10^{-4}$), as 
already done in Sec.~\ref{sec:results1} and study the spatial dependence of the correlation
functions (\ref{WeeksCorrel}).
We observe that the correlations in mobility measured by $C_{\delta}$ increase as
the temperature is lowered, but the spatial extent remains within a few particle diameters.
Even shorter ranged are the transverse correlations $C_{\perp}$.
The longitudinal correlations $C_{\|}$, however, increase significantly with decreasing
temperatures as shown in Fig.~\ref{Cpara.fig}.
For small separations $r$, the correlation functions $C_a$ oscillate due to the short-range
ordering as measured by the pair correlation function.
For larger separations ($r\gtrsim 4$), the oscillations have decayed and we observe an exponential
decrease $C_{\|}(r)\propto e^{-r/\xi_{\|}}$.
We find the correlation length $\xi_{\|}$ to increase from $\xi_{\|}\approx 2.5$ at high temperatures
to almost $4$ 
upon approaching the glass transition.
Similar observations have been made in Ref.\ \cite{Chay_nonaffine}
in simulations of the NAD in amorphous solids
and in Ref.\ \cite{Weeks_correlatedmotion} for
displacements in hard sphere colloidal systems when increasing the particle concentration.
In the latter, however, the length scale was found to remain on the order of $3$ particle diameters,
increasing significantly only beyond the glass transition.
It was argued in Ref.\ \cite{Weeks_correlatedmotion}
that the long-range correlations of the longitudinal displacements
reflect the string-like cooperative motion observed in computer simulations
\cite{Donati_strings}.
If this is indeed the case, our findings suggest that these string-like motions  
arise due to underlying string-like rearrangements between nearby inherent structures.
At larger distances, one would expect a hydrodynamic or elastic-like response
where the correlations decay not exponentially but $\propto 1/r$ in three dimensions
\cite{DiDonna}.
In agreement with Ref.\ \cite{Weeks_correlatedmotion}, we find no indication of such a
behavior, probably because the correlation function has already decayed to such 
small values that the numerical data do not allow us to detect the expected power-law.

\section{Coarse-grained NAD}\label{sec:coarsegr}
We investigate correlations in the direction of the NAD vectors
with the help of a coarse-grained displacement field
around every particle,
\begin{equation} \label{dCG}
\Mism_j(b) = N_j^{-1}\sum_{k}\hat{\mismB}_kw_b(r_{jk}^{\rm dq}).
\end{equation}
Here, we have defined the orientations
$\hat{\mismB}_j=\mismB_j/|\mismB_j|$ and
$r_{jk}^{\rm dq}$ is the distance between particles $j$ and $k$ in the
inherent structure of deformed configuration.  
$X^{\rm dq}$, $N_j = \sum_{k}w_b(r_{jk}^{\rm dq})$ is
the number of neighbours of particles $j$ within a distance $b$, and
$w_b(r)=1$ if $r\leq b$ and zero else.
Since the mean length of the NADs is strongly temperature-dependent
(see Fig.~\ref{dlength_T.fig}),
we use the normalized displacements in the definition (\ref{dCG}) in
order to separate this aspect and focus instead on the correlations between vector orientations. 

\begin{figure}[h]
\vskip 0.5cm
\includegraphics[width=1.0\linewidth]{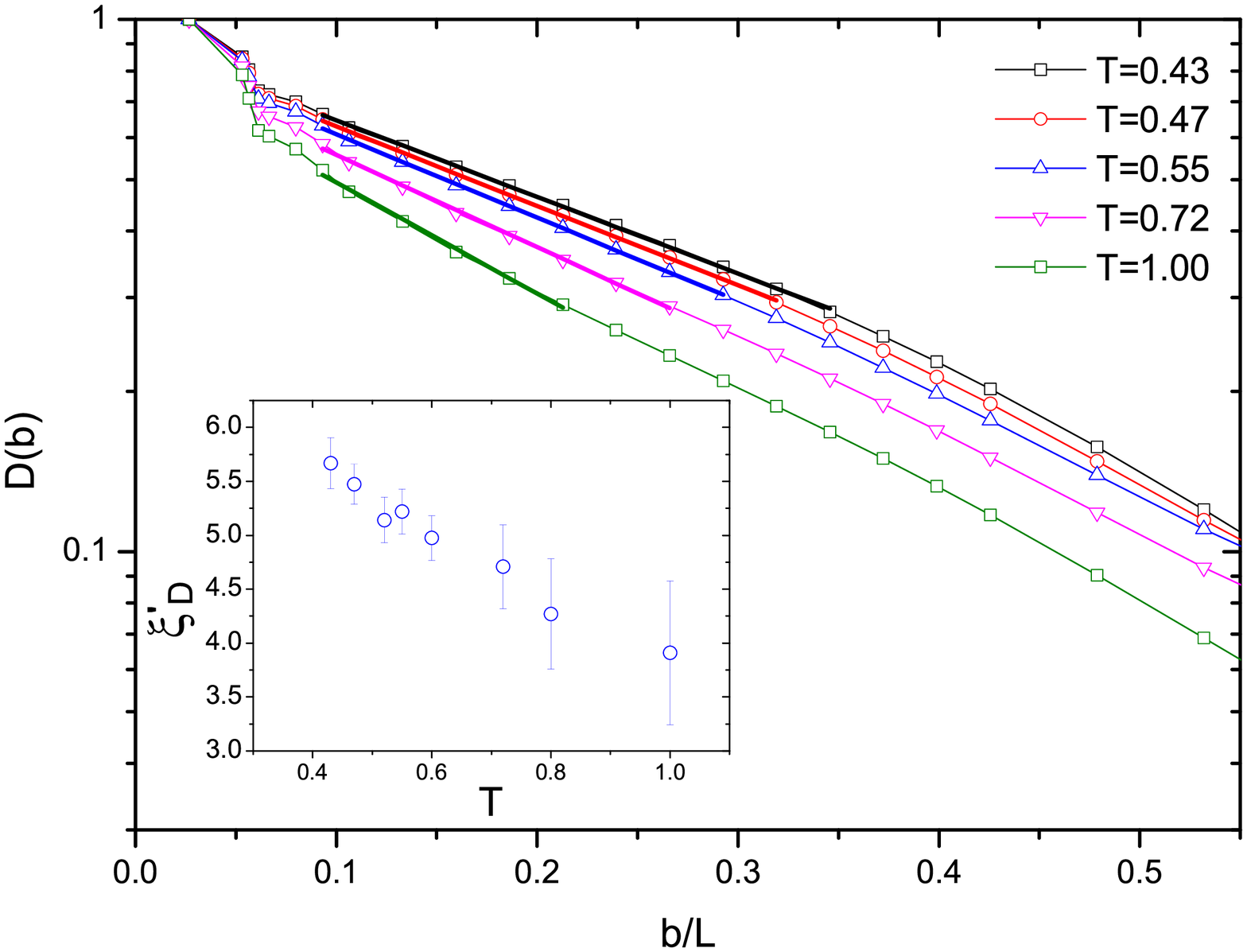}\\
  \caption{$\ave{\Mism^2}^{1/2}$ defined in Eq.~(\ref{dCG}) as function of $b$ for selected temperatures $T$.
  Results are shown for the BMLJ model under homogeneous shear with $\gamma=10^{-4}$.
  Inset shows the growing length scale obtained by fitting an exponential decaying function to data points which are shown in main figure.}
  \label{D_b.fig}
\vskip 0.5cm
\end{figure}

When the coarse-graining length $b$ is smaller than interparticle distances,
only one particle contributes to the average
in Eq.\ (\ref{dCG}) and $\Mism_j=\hat{\mismB}_j$.
As $b$ increases, more and more particles are involved in the average and
the magnitude of $\Mism_j$ decreases.
Figure \ref{D_b.fig} shows the function $D(b)=\ave{\Mism^2(b)}^{1/2}$
obtained for the BMLJ model when subjected to
homogeneous shear deformation with amplitude $\gamma=10^{-4}$.
The expected decrease of $D$ with $b$ is indeed observed.
For a fixed distance $b$, $D(b)$ increases monotonically with decreasing temperature.
This behavior clearly indicates increasing correlations between particle's
NAD orientations as the temperature is lowered.

Moreover, Fig.~\ref{D_b.fig} shows that these growing correlations extend
over distances larger than the particles diameter.
Since $D(b)$ measures the root-mean-square NAD direction when averaged over a length $b$,
we expect $D$ to decay when $b$ exceeds the size of a correlated region.
Therefore, the decay $D(b)\approx \exp{[-b/ \xiD]}$ allows to define a length
scale $\xiD$ over which the NAD vectors are correlated \cite{Fabien_prl}. 
At low temperatures, we find that $D(b)$ is well described by an exponential decay
over a wide range of smoothing lengths.
The length scale $\xiD$ that we obtain from a least-square fit to $D(b)$ is
shown in the inset of Fig.~\ref{D_b.fig} as a function of temperature.
At high temperatures ($T\gtrsim 0.8$), an exponential decay of $D(b)$ can be
observed only in a narrow interval $2\lesssim b \lesssim 4$.
In this regime, we find an approximately
temperature-independent length scale of $\xiD\approx 3$. This length scale
is somewhat larger than
the size of correlated liquid structure measured by the first peaks
in the pair correlation function.
Since local rearrangements have to include neighboring particles,
it is not surprising that $\xiD$ is on the order of a few particle diameter in this regime.
For decreasing temperatures, we observe an increase of the correlation
length $\xiD$ to values of about $6$ particle diameters.
Qualitatively, the same observations are made also for the BMSS model.

\begin{figure}[h]
\vskip 0.5cm
\includegraphics[width=1.0\linewidth]{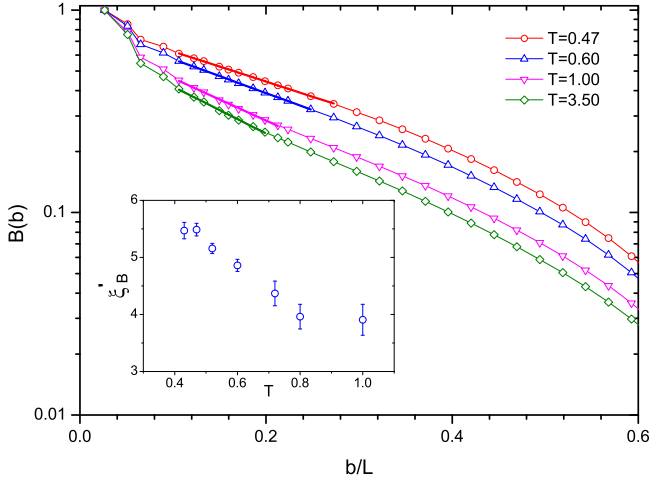}\\
  \caption{The coarse-graining function $B(b)$ defined in Eq.~(\ref{BarratB}) for selected temperatures $T$.
  Results are shown for the BMLJ model under homogeneous shear with $\gamma=10^{-4}$.
  Inset shows the growing length scale obtained by fitting an exponential decaying function to data points which are shown in main figure.}
  \label{B_b.fig}
\vskip 0.5cm
\end{figure}

In order to quantify the correlations in the NAD vectors in a different way
and to show that considering the magnitude in addition to the orientation
does not alter the qualitative behavior, 
we use the same function proposed in Ref.~\cite{Fabien_prl}
to study characteristic length scales of glasses,
\begin{equation} \label{BarratB}
B(b) = \ave{\sum_j{\bf U}_j(b)^2}^{1/2}/\avemism.
\end{equation}
In Eq.~(\ref{BarratB}), ${\bf U}_j(b)$ denotes the coarse-grained NAD vector
${\bf U}_j(b)=N_j^{-1}\sum_k \bmism_k w_b(r_{jk}^{\rm dq})$,
which is the analogue to Eq.\ (\ref{dCG}) for the NAD $\bmism_k$ instead of their unit vectors.
By construction, the function $B(b)$ has very similar limiting behavior as $D(b)$.
For large $b$, $B$ vanishes since $\sum_{k=1}^N\bmism_k$ corresponds to a
rigid translation of the whole system and is conventionally set to zero.
For small $b$, ${\bf U}_j\to \bmism_j$ and $B$ approaches one.
Figure \ref{B_b.fig} shows the function $B(b)$ for the BMLJ model subject to
homogeneous deformation with $\gamma=10^{-4}$.
We observe a very similar behavior of $B(b)$ compared to $D(b)$.
Thus, the qualitative conclusions are robust and hold for different definitions
of coarse-graining functions.
For a more quantitative comparison, we extract the correlation length $\xiB$ analogous to $\xiD$ above from
the decay $B(b)\approx\exp{[-b/ \xiB]}$
(see also the supplementary information of Ref.\ \cite{Majid_criticallength}. 
The growing length scale,
extracted by least-square fitting an exponentially decaying function to the data
is plotted in the inset of Fig.~\ref{B_b.fig}.
The length scales $\xiD$ and $\xiB$ not only show a very similar temperature-dependence
but are also quantitatively quite similar.

\subsection{Static length scale from histogram of coarse-grained NAD}
In addition to the quantities $D(b)$ and $B(b)$ that are averages of the
coarse-grained NAD, we have also studied the distribution of the coarse-grained 
NAD orientations $h_b(\Mism^2)$ and extracted a correlation length $\xi_{B}$ directly from them.
This correlation length displays the same type of temperature dependence of $\xi'_{B}$ and $\xi'_{D}$ 
and it is also quantitatively consistent with them. 

For small coarse-graining distances $b$, only a single particle contributes to
the sum in Eq.\ (\ref{dCG}).
Hence the histogram of $\Mism^2$ values is peaked around $1$, $h_b(\Mism^2)\to\delta(\Mism^2-1)$.
When coarse-graining over larger and larger distances $b$, $\Mism_j$ successively decreases and
therefore $h_b$ accumulates more weight at small values of $\Mism^2$
until $h_b$ becomes strongly peaked near $\Mism^2=0$ when $b$ becomes large
\cite{Majid_criticallength}.
The observed behavior bears some similarities to distribution of order parameters
near phase transitions, where the histogram switches between mostly ordered
to mostly disordered states when passing through the transition.  
Here, the transition between mostly correlated (peak of $h_b$ near $1$)
and uncorrelated (peak near $0$) regions happens at a value of $b$
which strongly increases with decreasing temperature.
Therefore, the histograms $h_b(\Mism^2)$ allow us to define a static correlation length $\xih$
as the value of $b$ that characterizes this transition.
As in Ref.\ \cite{Majid_criticallength}, we choose as definition
the coarse-graining distance $b$ for which the peak-location has shifted from one to $1/e$.
The length $\xih$ measures the average domain size
of NADs over which displacements are correlated. The temperature dependence of
$\xih$ for BMLJ (BMSS) model is shown in the inset of Fig.~\ref{xi_BMLJ.fig} (Fig.~\ref{xi_BMSS.fig}).
While the value of $\xih$ is on the order of a particle diameter
$\sigma_0$ at high temperature, it increases considerably by
cooling the system towards supercooled regions in quantitative agreement with the behavior of 
$\xi'_{B}$ and $\xi'_{D}$ discussed in Sec.~\ref{sec:coarsegr}
(see also $T$ values listed on Table~\ref{table:models}).
Alternatively one also can define $\xih$ as the coarse-graining distance
for which the variance $\ave{\Mism^4}-\ave{\Mism^2}^2$ is maximized. The
temperature dependence of $\xih$ is robust and does not depend on 
different definitions \cite{Majid_criticallength, Majid_thesis}.

How sensitive is the length scale on the amplitude $\gamma$ of
the applied shear deformation?
Figure \ref{xi_gamma.fig} shows that for small deformations,
$\gamma\lesssim 10^{-4}$, the correlation length $\xih$
becomes independent of $\gamma$.
This finding allows us to interpret
$\xih$ found above as an intrinsic property of the system,
characterizing the correlation between two nearby inherent structure configurations.
When increasing the deformation amplitude beyond $10^{-3}$, $\xih$ decreases
until the temperature-dependence is washed out by the shear.
In this regime, the shear deformation is strong enough to decorrelate the
initial and final inherent structure configuration,  
which may be interpreted in terms of meta-basin transitions \cite{ashwin}.

We want to stress once more that the procedure for obtaining the correlation length in this work is
performed just on simulation snapshots with static deformation. Hence our length scale has
a purely static character and its significant increase upon cooling can be
interpreted as a structural signature of different dynamical
regimes in supercooled liquids.

\subsection{Finite-size scaling analysis}
\begin{figure}
\vspace*{0.5cm}
\includegraphics[width=1.0\linewidth]{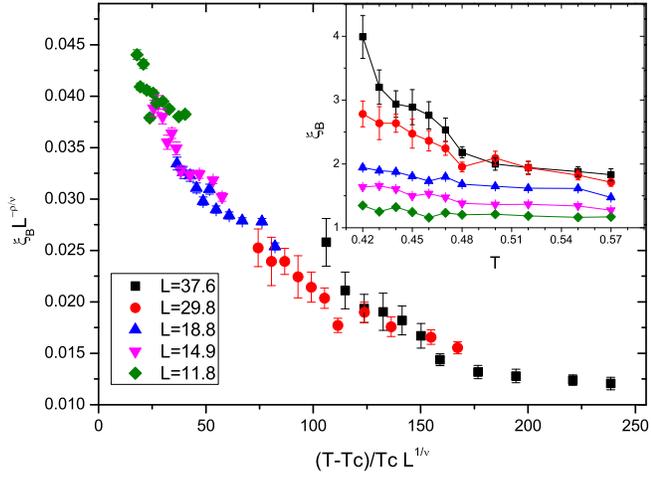}\\
\caption{Data collapse for the BMLJ model. Inset: Static correlation length $\xi_B$, extracted from histogram of coarse-grained NAD orientations as function of temperature for different system sizes.}
\label{xi_BMLJ.fig}
\end{figure}

\begin{figure}
\vspace*{0.5cm}
\includegraphics[width=1.0\linewidth]{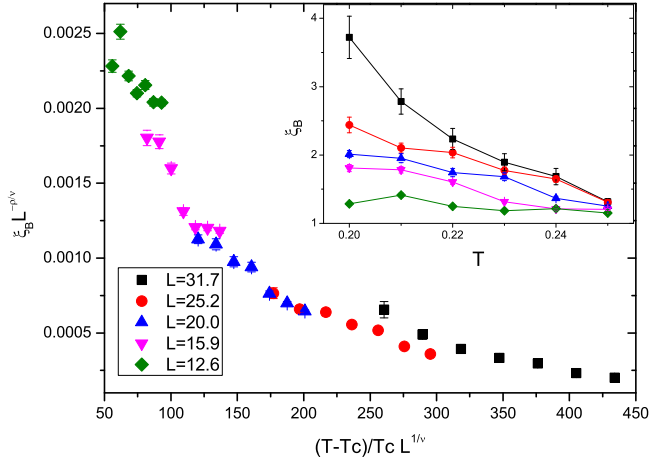}\\
\caption{Same as Fig.~\ref{xi_BMLJ.fig} but for the BMSS model.}
\label{xi_BMSS.fig}
\end{figure}

Figs.~\ref{xi_BMLJ.fig} and \ref{xi_BMSS.fig} show that $\xih$ depends strongly on the linear size $L$ of the system.
The strong system-size dependence together with the significant increase of $\xih$
is reminiscent of critical phenomena.
There, the growing correlation length of the order parameter characterizing
the transition gets cut-off by the system size.
To test if the system-size dependence of $\xih$ has any critical character
at low temperatures, we perform a finite-size scaling analysis.
We assume that the correlation length - associated with an unknown order parameter for glasses -
diverges with an exponent $\nu$ at the critical temperature $T_c$  
as $\xi\sim t^{-\nu}$, where $t$ is the reduced distance from the critical point, $t=(T-T_c)/T_c$.
Therefore, $\xih$ should diverge as well and in an infinite system $\xi_B\sim t^{-\rho}$,
where $\rho$ is the critical exponent characterizing its divergence.
On the basis of the scaling hypothesis for critical phenomena \cite{Barber},
the corresponding quantity $\xihL(T)$ in a finite system of
linear size $L$ should follow the behavior
\begin{equation} \label{scaling}
\xihL(T) \sim L^{\rho/\nu} Q_{\xih}(L^{1/\nu}t),
\end{equation}
where $Q_{\xih}(x)$ is a universal scaling function.
Therefore, plotting $\xihL(T)L^{-\rho/\nu}$ as a function of the
scaling variable $tL^{1/\nu}$ should collapse all data points for
different system sizes onto one master curve. 
The occurrence of a thermodynamic phase transition 
has been put forward by 
Random First Order Theory (RFOT) and related theories of the glass 
transition \cite{wolynes89,cavagna_pedestrian} and it has been discussed whether
this could take place at the temperature where the extrapolated configurational 
entropy vanishes $T_{\rm K}$ (the Kauzmann temperature). 
Fig.~\ref{xi_BMLJ.fig} shows the data collapse obtained for the the BMLJ model 
using $T_{\rm K}$ as the critical temperature 
\cite{wolynes89,cavagna_pedestrian}. 
For the BMLJ model, $T_{\rm K}$ is estimated numerically 
to be $T_{\rm K}\approx 0.30$ \cite{Sciortino_IS99,Coluzzi00}.
Fixing this value for $T_c$, we have varied the critical exponents $\nu$ and $\rho$.
The best data collapse is obtained for the values $\rho\approx 0.9 \pm 0.1$ and
$\nu\approx 0.65 \pm 0.1$, which is the case shown in Fig.~\ref{xi_BMLJ.fig}.
The error bars for the critical exponents are estimated
by varying the latter until the quality of data collapse starts to get worse.
Because of very long equilibration time of the system at low temperature 
and of the small increase of $\xi_B$ for small system sizes, our data are still
relatively far from the hypothetical critical point and this might be one of 
the reasons for the rather large uncertainties in the values of these 
exponents.
The good collapse of the numerical data, however, indicates that the critical region is
large enough to be detected in the temperature range where we could equilibrate our systems.
In Ref.~\cite{Majid_criticallength}, we have performed the finite-size scaling analysis by lifting 
the assumption $T_c=T_{\rm K}$. The results showed that one still could get a data collapse for 
values of $T_c$ which lay within the interval $0.25 \leqslant Tc \leqslant 0.4$.
For the BMSS model, $T_{\rm K}$ is estimated to be $T_{\rm K}\approx 0.11$ \cite{Coluzzi99}.
Fixing $T_c$ to this value, the best data collapse is obtained
by choosing the critical exponents as $\rho\approx 1.5 \pm 0.2$ and $\nu\approx 0.60 \pm 0.15$.
The case is shown in Fig.~\ref{xi_BMSS.fig}.
The value of the exponent $\nu$ is similar for BMLJ and BMSS models and this suggests
that the divergence of the underlying correlation length $\xi$ associated 
with the order parameter of the transition has the same origin in both cases. 
In Ref.~\cite{Majid_criticallength}, we have shown that also 
the dependence of relaxation time and configurational entropy 
on $\xi_B$ for the BMLJ model are in good agreement with predictions of RFOT.
It is interesting to notice that, within the standard RFOT \cite{wolynes89, 
BouchaudBiroli_mosaic}, the mosaic length scale $\xi_m$ is expected to grow 
significantly only at very low temperatures $(T<T_{\rm MCT})$ where long-lived 
metastable states occur, whereas our results indicate that the characteristic 
length scale $\xi_B$ starts to grow already at relatively higher temperatures. 
The growth of the so-called point-to-set correlation length 
at $(T>T_{\rm MCT})$  has also been observed in recent numerical simulations 
of three glass formers in Ref.~\cite{Hocky}. 
Overall, the picture emerging from our results is qualitatively consistent with the 
RFOT scenario. On the other hand, the uncertainties in the values of the critical 
exponents, as well as different results obtained in other recent studies \cite{Tanaka_critical}, 
indicate that further quantitative investigations are still needed to better clarify the nature of 
the critical behavior observed here.

\section{Conclusions} \label{sec:concl} 

Cooperatively rearranging regions (CRR) play an important role for the dynamics of
supercooled liquids with their sizes growing moderately upon approaching
the glass transition \cite{Glotzer_length99,Ludo_Science05,Weeks_correlatedmotion}.
Very recently, structural signatures of the CRR have been detected, suggesting
medium-range order \cite{Tanaka_DHandPsi6} or localized soft modes \cite{WidmerCooperSoftModes} 
as triggers of the rearrangements.
Here, we have shown by extensive molecular simulations that correlations in 
neighboring IS are quite reminiscent of CRR that are observed 
in the system's dynamics \cite{Weeks_correlatedmotion}.
The behavior we report here is found for different models of
fragile glasses, i.e., the Kob-Andersen model and a binary soft sphere mixture.
The distance between two IS -- that are related via shear deformation  of rather
small amplitude $\gamma$ -- sharply decreases with decreasing temperature
below the onset temperature of the landscape-dominated regime \cite{Sastry_TainKA}.
We observe that the mean distance between the two IS varies exponentially
with the inherent structure energy $e_{\rm IS}$.  
Furthermore, we observe a crossover between two regimes that is also present
in a qualitative change of the NAD distribution from an exponential to
a power-law shape.
The exponent of the latter can be rationalized from elasticity arguments \cite{Dyre99}.
We use different measures for correlations of NAD vectors.
Qualitatively, we find similar results as
experiments on correlated motions in colloids \cite{Weeks_correlatedmotion}.
Quantitatively, the static correlation length extracted from NAD of IS 
grows significantly upon getting closer to the glass transition. Moreover, the finite-size scaling 
of our results indicates that the correlation length diverges at low temperatures.
Since thermal fluctuations tend to wash out the correlations in the NAD, 
our method is very sensitive and provides an efficient tool to investigate 
correlated regions. This is a first interesting outcome of this work. 
It is worth noting, with this respect, that NAD field has proven to be an insightful 
investigation tool also for the mechanics of amorphous solids \cite{Fabien_prl} 
and hence the approach we propose has a good potential to bridge the investigation of supercooled 
liquids to the one of amorphous solids, within a unifying picture of glass transition. In addition, the results here 
discussed suggest that the long range spatial correlations detected by the NAD field upon lowering the temperature might 
be indeed the spatial correlations underlying the onset and development of cooperative dynamics typically observed in 
supercooled liquids. This would be a significant new insight into the physics of supercooled liquids and more work is 
currently in progress to clarify this issue. The critical growth of spatial correlations that we have reported
here overall supports the glass transition scenario based on Random First Order Theory and the analysis done 
seems to be quite robust. On the other hand,  
the finite-size scaling proposed also raises the question of identifying the critical 
temperature and for the moment  we cannot rule out different possibilities: this requires further and more quantitative 
analysis. Finally, it would be extremely interesting, at this point, to be able to directly relate the 
growth of the correlation length and the features of the NAD field to the change in the transport 
properties of the system, i.e. in its viscoelastic response: whereas, to some extent, we can 
relate our results to the viscosity increase through the RFOT \cite{Majid_criticallength}, 
a more coherent connection with the mechanical response and with the onset of 
rigidity in the material needs to be developed. 

\subsection*{Acknowledgements}
We thank Ludovic Berthier, Andrea Cavagna, Walter Kob, Srikanth Sastry and Anne Tanguy for useful discussions. 
Computational resources of the PolyHub Virtual Organization is greatly acknowledged.
EDG is supported by the Swiss National Science Foundation (SNSF Grant No. PP002 126483/1).

%


\newpage


\end{document}